\documentclass[a4paper,aps,twocolumn,nofootinbib]{revtex4}
\RequirePackage[colorlinks,hyperindex]{hyperref}
\RequirePackage[english]{babel}
\RequirePackage[latin1]{inputenc}
\RequirePackage[T1]{fontenc}
\RequirePackage{mathrsfs}
\RequirePackage{amsmath}
\RequirePackage{amssymb}
\RequirePackage{amsbsy}
\RequirePackage{color}
\RequirePackage{bm}
\hypersetup{colorlinks=true,breaklinks=true,urlcolor=blue,linkcolor=red}
\begin{document}
\title{\bf{Euler and Pontryagin currents of the Dirac operator}}
\author{Luca Fabbri$^{\nabla}$\!\!\! $^{\hbar}$\footnote{luca.fabbri@unige.it}}
\affiliation{$^{\nabla}$DIME, Universit\`{a} di Genova, Via all'Opera Pia 15, 16145 Genova, ITALY\\
$^{\hbar}$INFN, Sezione di Genova, Via Dodecaneso 33, 16146 Genova, ITALY}
\date{\today}
\begin{abstract}
On differential manifolds with spinor structure, it is possible to express the Euler and Pontryagin currents in terms of tensors that also appear as source in the Dirac equation. It is hence possible to tie concepts rooted in geometry and topology to dynamical characters of quantum matter.
\end{abstract}
\maketitle
\section{Introduction}
In differential geometry, the curvature of a Riemannian manifold encodes the properties of the space. Among all scalars that are formed from the curvature, some can be expressed as the divergence of a suitable vector, and thus they describe topological features of that space, which are the characteristic classes. For example, in $2$-dimensional spaces, the characteristic class is the Euler class $\chi_{2}$ while, in $4$-dimensional cases, they are the Euler and Pontryagin classes $\chi_{4}$ and $p_{4}$ respectively. Similar considerations are true also for Riemann-Cartan manifolds, where curvature is accompanied by torsion. In such a case, the Pontryagin class is called Nieh-Yan class \cite{Chandia:1997hu, Camargo:2022gcw, Nascimento:2021vou}. For electrodynamics, the Pontryagin class is $F_{\mu\nu}F_{\rho\sigma}\varepsilon^{\mu\nu\rho\sigma}$ as very well known.

The vector whose divergence is the characteristic class is called topological current and it can be computed with straightforward manipulations. In electrodynamics, it is given by $\varepsilon^{\rho\mu\nu\sigma}F_{\mu\nu}A_{\sigma}$ as easy to check. On torsional manifolds, it is the Hodge dual of the torsion tensor \cite{Nascimento:2021vou}. And for Riemannian manifolds, it is given by a suitable combination of spin connections, and therefore by frames.

In this last case, for our purpose it is not important to give the topological current explicitly in terms of frames, as it is enough to state that, for the manner in which it is constructed, it behaves like a true vector only when taken in divergences. The same for topological currents in electrodynamics, where the vector $\varepsilon^{\rho\mu\nu\sigma}F_{\mu\nu}A_{\sigma}$ is not gauge invariant, and it is only when it is taken in a divergence that the spurious term $\nabla_{\rho}(\varepsilon^{\rho\mu\nu\sigma}F_{\mu\nu}\nabla_{\sigma}\varphi)$ is identically equal to zero, ensuring the gauge invariance of the topological current. Yet, it may be important to have a way to express the topological currents in terms of true vectors.

Because the characteristic classes are given in terms of curvatures, this problem can be reduced to that of asking whether it is possible to write curvatures in terms of true tensors, and not just connections. In this work we shall see that when the manifold possesses a spin structure, it has enough richness to acquire the possibility to convert spin connection and gauge potential into a real tensor and a gauge invariant vector in terms of which the Riemann curvature and the Maxwell strength can be expressed \cite{Fabbri:2021mfc}.

This will enables us to express also the topological currents in terms of real vectors. In addition, we will see that the vectors in terms of which the topological currents are expressed also enter as sources into the Dirac field equations, henceforth establishing a link between topological objects and dynamical effects on the Dirac operator.

To help seeing these concepts, we will furnish some examples for different spinorial systems in low-dimensional cases, that is for dimensions $n\!=\!2,\ 3,\ 4$ \cite{Abanov:1999qz, Golkar:2014paa}.
\section{General preliminaries}\label{II}
\subsection{Geometrical and Topological notions}
Let it be given a manifold $\mathscr{M}$ with metric $g_{\sigma\nu}$ and with frame and co-frame $e^{a}_{\sigma}$ and $e^{\alpha}_{s}$ such that
\begin{eqnarray}
e^{i}_{\sigma}e^{\alpha}_{i}\!=\!\delta^{\alpha}_{\sigma}
\ \ \ \ \ \ \ \ e^{a}_{\nu}e^{\nu}_{s}\!=\!\delta^{a}_{s}
\end{eqnarray}
verifying
\begin{eqnarray}
e_{i}^{\sigma}e_{j}^{\alpha}g_{\sigma\alpha}\!=\!\eta_{ij}
\end{eqnarray}
with $\eta_{ij}$ the Minkowskian matrix. This matrix is diagonal with elements that are all unitary up to the sign, and with it we will specify dimension and signature. Greek indices are coordinate indices on the manifold, transforming with diffeomorphisms, while Latin indices are world indices on the tangent space, transforming with elements of the real Lorentz group. With the metric $g_{\sigma\nu}$ and its inverse $g^{\sigma\nu}$ we lower/raise coordinate indices, with the Minkowskian metric and its inverse we lower/raise world indices, while with frame and co-frame we convert coordinate to world indices. This is commonly known in differential geometry although interested readers may find more details in \cite{G}.

The spin connection can be introduced by the relation
\begin{eqnarray}
C^{i}_{\phantom{i}k\mu}\!=\!e^{i}_{\sigma}\partial_{\mu}e_{k}^{\sigma}
\!+\!e^{i}_{\alpha}e_{k}^{\sigma}\Lambda^{\alpha}_{\sigma\mu}
\label{connection}
\end{eqnarray}
where $\Lambda^{\alpha}_{\sigma\mu}$ is the Levi-Civita connection, entirely written in terms of the derivatives of the metric. See again \cite{G}.

The Riemann curvature of the Levi-Civita connection can be converted into $R_{ij\mu\nu}\!=\!e^{\rho}_{i}e^{\sigma}_{j}R_{\rho\sigma\mu\nu}$ so that
\begin{eqnarray}
&R^{i}_{\phantom{i}j\mu\nu}\!=\!\partial_{\mu}C^{i}_{\phantom{i}j\nu}
\!-\!\partial_{\nu}C^{i}_{\phantom{i}j\mu}
\!+\!C^{i}_{\phantom{i}k\mu}C^{k}_{\phantom{k}j\nu}
\!-\!C^{i}_{\phantom{i}k\nu}C^{k}_{\phantom{k}j\mu}
\end{eqnarray}
as is straightforward to demonstrate \cite{G}.

From it we define $R^{i}_{\phantom{i}j\mu\nu}e^{\mu}_{i}\!=\!R_{j\nu}$ called Ricci curvature and hence we can define $R_{a\nu}e^{\nu}_{c}\eta^{ac}\!=\!R$ called Ricci scalar, which is a topological invariant in $2$ dimensions. For more scalars we have to consider products of two curvatures as $R_{\rho\sigma\mu\nu}R^{\rho\sigma\mu\nu}$ or $R_{\mu\nu}R^{\mu\nu}$ beside the obvious $R^{2}$ amounting to all non-trivial independent contractions, and for which the Gauss-Bonnet term $R_{\rho\sigma\mu\nu}R^{\rho\sigma\mu\nu}\!-\!4R_{\mu\nu}R^{\mu\nu}\!+\!R^{2}$ is a topological invariant in $4$ dimensions. In $4$ dimensions it is possible to write the Gauss-Bonnet term like
\begin{eqnarray}
\nonumber
&R^{\alpha\nu\mu\rho}R_{\rho\mu\nu\alpha}\!-\!4R^{\alpha\mu}R_{\mu\alpha}\!+\!R^{2}\equiv\\
&\equiv-\frac{1}{4}\varepsilon^{\alpha\sigma\pi\tau}\varepsilon^{\rho\omega\mu\nu}
R_{\alpha\sigma\mu\nu}R_{\pi\tau\rho\omega}
\end{eqnarray}
and it is also possible to have another topological invariant given by $\frac{1}{4}\varepsilon^{\mu\nu\pi\eta}R^{\alpha\sigma}_{\phantom{\alpha\sigma}\mu\nu}R_{\alpha\sigma\pi\eta}$ and which is parity-odd.

The three above topological invariants characterize the Euler classes $\chi_{2}$ and $\chi_{4}$ and the Pontryagin class $p_{4}$ respectively. As such we must have that
\begin{eqnarray}
&\chi_{2}:\ \ \longrightarrow\ \ \ \ -\frac{1}{2}R\!=\!\nabla_{\rho}G^{\rho}_{2}
\end{eqnarray}
and
\begin{eqnarray}
&\chi_{4}:\ \ \longrightarrow\ \ \ \ -\frac{1}{8}\varepsilon^{\alpha\sigma\pi\tau}\varepsilon^{\rho\omega\mu\nu}
R_{\alpha\sigma\mu\nu}R_{\pi\tau\rho\omega}\!=\!\nabla_{\rho}G^{\rho}_{4}
\end{eqnarray}
\begin{eqnarray}
&p_{4}:\ \ \longrightarrow\ \ \ \ \frac{1}{4}\varepsilon^{\mu\nu\pi\eta}R^{\alpha\sigma}_{\phantom{\alpha\sigma}\mu\nu}R_{\alpha\sigma\pi\eta}\!=\!\nabla_{\rho}K^{\rho}_{4}
\end{eqnarray}
for some bi-dimensional vector $G^{\rho}_{2}$ and four-dimensional vector $G^{\rho}_{4}$ and axial-vector $K^{\rho}_{4}$ (the fractions in front of the curvatures are for later convenience). Further details on characteristic classes can be found in reference \cite{DFN}.
\subsection{Quantum fields}
Let it be given on $\mathscr{M}$ a spinorial structure determined by the Clifford algebra $\{\boldsymbol{\gamma}^{a},\boldsymbol{\gamma}^{b}\}\!=\!2\mathbb{I}\eta^{ab}$ and by a spin-$1/2$ spinor field. We define $\left[\boldsymbol{\gamma}_{a},\!\boldsymbol{\gamma}_{b}\right]/4\!=\!\boldsymbol{\sigma}_{ab}$ as the elements of the complex Lorentz algebra and their exponentiation are the elements of the complex Lorentz group with which a spin-$1/2$ spinor field transforms. When complex Lorentz transformations are accompanied by a phase transformation we talk about spinorial transformations \cite{G}.

We introduce the spinorial connection according to
\begin{eqnarray}
&\boldsymbol{C}_{\mu}\!=\!\frac{1}{2}C_{ij\mu}\boldsymbol{\sigma}^{ij}\!+\!iqA_{\mu}\mathbb{I}
\label{conn}
\end{eqnarray}
in terms of the spin connection $C_{ij\mu}$ and an abelian gauge potential $A_{\mu}$ having $q$ as charge. With it
\begin{eqnarray}
&\boldsymbol{\nabla}_{\mu}\psi\!=\!\partial_{\mu}\psi\!+\!\boldsymbol{C}_{\mu}\psi
\label{der}
\end{eqnarray}
is the covariant derivative of the spinor field.

As usual, we have that
\begin{eqnarray}
&[\boldsymbol{\nabla}_{\mu},\! \boldsymbol{\nabla}_{\nu}]\psi
\!=\!\frac{1}{2}R_{ij\mu\nu}\boldsymbol{\sigma}^{ij}\psi\!+\!iqF_{\mu\nu}\psi
\label{comm}
\end{eqnarray}
is a general identity.

The dynamics is assigned by the Dirac equation
\begin{eqnarray}
&i\boldsymbol{\gamma}^{\mu}\boldsymbol{\nabla}_{\mu}\psi\!-\!m\psi\!=\!0
\end{eqnarray}
in which we have assumed no torsion for simplicity.

This construction is valid for any dimension. However, for our purposes, we need to introduce the technique of polar decomposition, and this can be done only after that we specify dimension and signature of the space. In what follows, we will consider spaces of dimension $4$ and lower, starting from the four-dimensional space since this is the one for which some techniques have already been developed \cite{Fabbri:2021mfc}, then proceeding with decreasing complexity.
\section{Polar Degrees of Freedom and Tensorial Connections}\label{III}
\subsection{4-dimensions}\label{A}
\subsubsection{(1+3)-signature}
We begin with the physical space. In it, the Minkowski matrix has four elements, of which one is equal to unity and three are equal to minus unity, while the Levi-Civita completely antisymmetric tensor $\varepsilon_{\alpha\nu\sigma\tau}$ has four indices.

The Clifford matrices, and in particular the sigma matrices, verify $2i\boldsymbol{\sigma}_{ab}
\!=\!\varepsilon_{abij}\boldsymbol{\sigma}^{ij}\boldsymbol{\pi}$ implicitly defining an additional matrix $\boldsymbol{\pi}$ which is parity-odd.\footnote{This matrix is usually designated as a gamma with an index five, but we will not employ this notation here: the index five simply has no sense, especially in the three- and two-dimensional cases we will consider in the following. In addition, we will indicate it with the boldface Greek letter $\boldsymbol{\pi}$, whose Latin correspondent is $p$, to mark the fact that it is \emph{parity}-odd (much in the same way in which the generators are denoted with the boldface Greek letter $\boldsymbol{\sigma}$, whose Latin is $s$, to indicate that they are \emph{spin}-dependent).} We have
\begin{eqnarray}
&\boldsymbol{\gamma}_{i}\boldsymbol{\gamma}_{j}\boldsymbol{\gamma}_{k}
\!=\!\boldsymbol{\gamma}_{i}\eta_{jk}
\!-\!\boldsymbol{\gamma}_{j}\eta_{ik}
\!+\!\boldsymbol{\gamma}_{k}\eta_{ij}
\!-\!i\varepsilon_{ijkq}\boldsymbol{\gamma}^{q}\boldsymbol{\pi}
\label{id4}
\end{eqnarray}
showing that products of more than three Clifford matrices can always be reduced to the product of two. Defining $\overline{\psi}\!=\!\psi^{\dagger}\boldsymbol{\gamma}^{0}$ we can build the bi-linear spinor quantities
\begin{eqnarray}
&S^{a}\!=\!\overline{\psi}\boldsymbol{\gamma}^{a}\boldsymbol{\pi}\psi\ \ \ \
\ \ \ \ \ \ \ \ U^{a}\!=\!\overline{\psi}\boldsymbol{\gamma}^{a}\psi\label{vectors}\\
&\Theta\!=\!i\overline{\psi}\boldsymbol{\pi}\psi\ \ \ \
\ \ \ \ \ \ \ \ \Phi\!=\!\overline{\psi}\psi\label{scalars}
\end{eqnarray}
which are all real tensors. With them, we have
\begin{eqnarray}
&2U_{\mu}S_{\nu}\boldsymbol{\sigma}^{\mu\nu}\boldsymbol{\pi}\psi
\!+\!U^{2}\psi=0\label{AUX}
\end{eqnarray}
and
\begin{eqnarray}
&U_{a}U^{a}\!=\!-S_{a}S^{a}\!=\!\Theta^{2}\!+\!\Phi^{2}\label{NORM}\\
&U_{a}S^{a}\!=\!0\label{ORTHOGONAL}
\end{eqnarray}
as three cases of the known Fierz identities \cite{PS, O, Nishi:2004st, Nieves:2003in, Dale:2022jub, HoffdaSilva:2019xvd, CoronadoVillalobos:2020yvr}. When $\Phi^{2}\!+\!\Theta^{2}\!\neq\!0$ it is always possible to write a spinor in polar form, which is given, in chiral representation, as
\begin{eqnarray}
&\psi\!=\!\phi\ e^{-\frac{i}{2}\beta\boldsymbol{\pi}}
\ \boldsymbol{L}^{-1}\left(\begin{tabular}{c}
$1$\\
$0$\\
$1$\\
$0$
\end{tabular}\right)
\label{spinor}
\end{eqnarray}
for a pair of functions $\phi$ and $\beta$ and for some $\boldsymbol{L}$ with the structure of spinor transformations \cite{jl1, jl2}. The bi-linear spinor scalars are then given by
\begin{eqnarray}
&\Theta\!=\!2\phi^{2}\sin{\beta}\ \ \ \ \ \
\ \ \ \ \ \ \Phi\!=\!2\phi^{2}\cos{\beta}
\end{eqnarray}
so that $\phi$ and $\beta$ are a scalar and a pseudo-scalar, called module and chiral angle. We can also define
\begin{eqnarray}
&S^{a}\!=\!2\phi^{2}s^{a}\ \ \ \
\ \ \ \ \ \ \ \ U^{a}\!=\!2\phi^{2}u^{a}
\end{eqnarray}
being the velocity vector and spin axial-vector. Identities (\ref{AUX}) and (\ref{NORM}-\ref{ORTHOGONAL}) reduce to
\begin{eqnarray}
&2u_{\mu}s_{\nu}\boldsymbol{\sigma}^{\mu\nu}\boldsymbol{\pi}\psi\!+\!\psi=0\label{aux}
\end{eqnarray}
and
\begin{eqnarray}
&u_{a}u^{a}\!=\!-s_{a}s^{a}\!=\!1\label{norm}\\
&u_{a}s^{a}\!=\!0\label{orthogonal}
\end{eqnarray}
showing that the velocity has only $3$ independent components (the $3$ components of its spatial part) whereas the spin has $2$ independent components (the $2$ angles that, in the rest-frame, its spatial part forms with one given axis, usually chosen as the third). Therefore $\boldsymbol{L}$ is the Lorentz transformation that takes any general spinor into its rest frame with spin aligned along the third axis. As for the sixth parameter of $\boldsymbol{L}$ it can be taken as the angle of the rotation around the third axis (or more in general, around the direction of the spin) or as a global phase (that is, a gauge phase). These two are indistinguishable for spinors in rest-frame and spin-eigenstate. The only $2$ remaining components that cannot be transferred into the frame are the $\phi$ and $\beta$ scalars. They are the degrees of freedom.

It is a general result that the logarithmic derivative of an element of a Lie group belong to its Lie algebra, or
\begin{eqnarray}
&\boldsymbol{L}^{-1}\partial_{\mu}\boldsymbol{L}\!=\!iq\partial_{\mu}\zeta\mathbb{I}
\!+\!\frac{1}{2}\partial_{\mu}\zeta_{ij}\boldsymbol{\sigma}^{ij}\label{spintrans}
\end{eqnarray}
for some $\zeta_{ij}$ and $\zeta$ \cite{Fabbri:2021mfc}. With these, we can define
\begin{eqnarray}
&R_{ij\mu}\!=\!\partial_{\mu}\zeta_{ij}\!-\!C_{ij\mu}\label{R}\\
&P_{\mu}\!=\!q(\partial_{\mu}\zeta\!-\!A_{\mu})\label{P}
\end{eqnarray}
which are shown to be a real tensor and a gauge invariant vector (for the proof see reference \cite{Fabbri:2022kfr}). To identify them separately, we shall name them space-time and gauge tensorial connections. We can also have them both collected together into the single object
\begin{eqnarray}
&\!\!\!\!\!\!\!\!R^{\alpha\nu}_{\phantom{\alpha\nu}\mu}\!-\!2P_{\mu}u_{\rho}s_{\pi}\varepsilon^{\rho\pi\alpha\nu}\!=\!
-\frac{1}{2}\varepsilon^{\alpha\nu\sigma\tau}M_{\sigma\tau\mu}\!=\!
\Sigma^{\alpha\nu}_{\phantom{\alpha\nu}\mu}
\label{Sigma4}
\end{eqnarray}
with $M_{\alpha\nu\mu}$ and $\Sigma_{\alpha\nu\mu}$ the Hodge dual of each other. For this single expression, we will simply talk about tensorial connection. Having compacted the two tensorial connections into a unique expression, we can treat in the same formal way the pure space-time case ($\Sigma_{ij\mu}=R_{ij\mu}$) as well as the pure gauge case ($\Sigma_{ij\mu}=-2P_{\mu}u^{a}s^{b}\varepsilon_{abij}$). With it
\begin{eqnarray}
&\boldsymbol{\nabla}_{\mu}\psi\!=\!(\nabla_{\mu}\ln{\phi}\mathbb{I}
\!-\!\frac{i}{2}\nabla_{\mu}\beta\boldsymbol{\pi}
\!-\!\frac{1}{2}\Sigma_{\alpha\nu\mu}\boldsymbol{\sigma}^{\alpha\nu})\psi
\label{decspinderSigma}
\end{eqnarray}
in which identities $2i\boldsymbol{\pi}\boldsymbol{\sigma}_{ab}\!=\!\varepsilon_{abcd}\boldsymbol{\sigma}^{cd}$ and \eqref{aux} have been used. Additionally, we notice that
\begin{eqnarray}
&\nabla_{\mu}s_{\nu}\!=\!s^{\alpha}\Sigma_{\alpha\nu\mu}
\ \ \ \ \ \ \ \ \nabla_{\mu}u_{\nu}\!=\!u^{\alpha}\Sigma_{\alpha\nu\mu}\label{su}
\end{eqnarray}
are also valid as general identities \cite{Fabbri:2023dgv}.

By employing these polar variables, we have
\begin{eqnarray}
&\!\!\!\!\!\!\!\!R^{i}_{\phantom{i}j\mu\nu}\!=\!-(\nabla_{\mu}R^{i}_{\phantom{i}j\nu}
\!-\!\nabla_{\nu}R^{i}_{\phantom{i}j\mu}\!+\!R^{i}_{\phantom{i}k\mu}R^{k}_{\phantom{k}j\nu}
\!-\!R^{i}_{\phantom{i}k\nu}R^{k}_{\phantom{k}j\mu})\label{Riemann}\\
&\!\!\!\!qF_{\mu\nu}\!=\!-(\nabla_{\mu}P_{\nu}\!-\!\nabla_{\nu}P_{\mu})\label{Maxwell}
\end{eqnarray}
showing that the space-time and gauge tensorial connections can be seen as the covariant potentials of Riemann curvature and Maxwell strength (the above relationships were proven in \cite{Fabbri:2022kfr}). They are also known as space-time and gauge curvatures. As before, then, we can have them collected into the single object given by
\begin{eqnarray}
&\!\!\!\!R^{ij}_{\phantom{ij}\mu\nu}\!-\!2qF_{\mu\nu}u_{a}s_{b}\varepsilon^{abij}\!=\!
-\frac{1}{2}\varepsilon^{ijab}M_{ab\mu\nu}\!=\!\Sigma^{ij}_{\phantom{ij}\mu\nu}\label{curvSigma}
\end{eqnarray}
with $M_{ab\mu\nu}$ and $\Sigma_{ab\mu\nu}$ being a particular Hodge dualization in two indices. A very straightforward check gives
\begin{eqnarray}
&\!\!\!\!\!\!\!\!\Sigma^{i}_{\phantom{i}j\mu\nu}\!=\!-(\nabla_{\mu}\Sigma^{i}_{\phantom{i}j\nu}
\!-\!\nabla_{\nu}\Sigma^{i}_{\phantom{i}j\mu}
\!+\!\Sigma^{i}_{\phantom{i}k\mu}\Sigma^{k}_{\phantom{k}j\nu}
\!-\!\Sigma^{i}_{\phantom{i}k\nu}\Sigma^{k}_{\phantom{k}j\mu})\label{curv}
\end{eqnarray}
which we call simply curvature. With it, (\ref{comm}) becomes
\begin{eqnarray}
&[\boldsymbol{\nabla}_{\mu},\! \boldsymbol{\nabla}_{\nu}]\psi
\!=\!\frac{1}{2}\Sigma_{ab\mu\nu}\boldsymbol{\sigma}^{ab}\psi
\label{commutator}
\end{eqnarray}
in its most compact form.

The cyclic permutation of commutators gives
\begin{eqnarray}
&\varepsilon^{\kappa\mu\nu\rho}\nabla_{\mu}\Sigma^{\eta\tau}_{\phantom{\eta\tau}\nu\rho}\!=\!0:
\label{JB}
\end{eqnarray}
the pure space-time case reduces to the Bianchi identity, while the pure gauge case reduces to the Cauchy identity.

Finally, the Dirac equation is equivalent to the pair
\begin{eqnarray}
&\nabla_{\mu}\beta\!+\!M_{\mu}\!+\!2ms_{\mu}\cos{\beta}\!=\!0\label{dep1}\\
&\nabla_{\mu}\ln{\phi^{2}}\!+\!\Sigma_{\mu}\!+\!2ms_{\mu}\sin{\beta}\!=\!0\label{dep2}
\end{eqnarray}
with $\Sigma^{\alpha}\!=\!\Sigma^{\alpha\nu}_{\phantom{\alpha\nu}\nu}$ and $M^{\alpha}\!=\!M^{\alpha\nu}_{\phantom{\alpha\nu}\nu}$ as sources \cite{Fabbri:2023onb}.

With the tensorial connection (\ref{Sigma4}) one can prove that the curvature (\ref{curvSigma}) verifies the following identities
\begin{eqnarray}
&-\frac{1}{4}\varepsilon^{\rho\omega\mu\nu}\Sigma_{\alpha\sigma\mu\nu}
M^{\alpha\sigma}_{\phantom{\alpha\sigma}\rho\omega}\!=\!\nabla_{\mu}G^{\mu}_{4}\label{divvecG4}\\
&\frac{1}{4}\varepsilon^{\mu\nu\pi\eta}\Sigma^{\alpha\sigma}_{\phantom{\alpha\sigma}\mu\nu}
\Sigma_{\alpha\sigma\pi\eta}\!=\!\nabla_{\mu}K^{\mu}_{4}\label{divvecK4}
\end{eqnarray}
for some vector and axial-vector
\begin{eqnarray}
&\!\!\!\!\!\!\!\!G^{\mu}_{4}\!=\!-\frac{1}{4}\varepsilon^{\mu\nu\eta\pi}\varepsilon^{\alpha\sigma\omega\rho}
\Sigma_{\omega\rho\nu}(\Sigma_{\sigma\alpha\eta\pi}
-\!\frac{2}{3}\Sigma_{\sigma\kappa\eta}\Sigma^{\kappa}_{\phantom{\kappa}\alpha\pi})\label{vecG4}\\
&\!\!\!\!K^{\mu}_{4}\!=\!\frac{1}{2}\varepsilon^{\mu\nu\eta\pi}\Sigma^{\alpha}_{\phantom{\alpha}\sigma\nu}
(\Sigma^{\sigma}_{\phantom{\sigma}\alpha\eta\pi}
\!-\!\frac{2}{3}\Sigma^{\sigma}_{\phantom{\sigma}\kappa\eta}
\Sigma^{\kappa}_{\phantom{\kappa}\alpha\pi})\label{vecK4}
\end{eqnarray}
written in terms of the tensorial connection alone: when no electrodynamics is present they reduce to
\begin{eqnarray}
&G^{\mu}_{4}\!=\!-\frac{1}{4}\varepsilon^{\mu\nu\eta\pi}\varepsilon^{\alpha\sigma\omega\rho}
R_{\omega\rho\nu}(R_{\sigma\alpha\eta\pi}
\!-\!\frac{2}{3}R_{\sigma\kappa\eta}R^{\kappa}_{\phantom{\kappa}\alpha\pi})\\
&K^{\mu}_{4}\!=\!\frac{1}{2}\varepsilon^{\mu\nu\eta\pi}R^{\alpha}_{\phantom{\alpha}\sigma\nu}
(R^{\sigma}_{\phantom{\sigma}\alpha\eta\pi}
\!-\!\frac{2}{3}R^{\sigma}_{\phantom{\sigma}\kappa\eta}
R^{\kappa}_{\phantom{\kappa}\alpha\pi});
\end{eqnarray}
when no curvature is present we have
\begin{eqnarray}
&G^{\mu}_{4}\!=\!0\\
&K^{\mu}_{4}\!=\!4qF_{\eta\pi}P_{\nu}\varepsilon^{\eta\pi\nu\mu}
\end{eqnarray}
which are the Euler and Pontryagin topological currents.

We have written the Euler and Pontryagin topological currents with space-time and gauge tensorial connections and we have seen that these tensorial connections enter as sources in the Dirac differential field equations.
\subsection{2-dimensions}
\subsubsection{(1+1)-signature}
The $(1\!+\!1)$-dimensional case has the Minkowski matrix with two elements of opposite sign, while the Levi-Civita completely antisymmetric tensor has only two indices.

The Clifford matrices, and in particular the sigma matrices, verify $2\boldsymbol{\sigma}_{ab}\!=\!\varepsilon_{ab}\boldsymbol{\pi}$ defining the $\boldsymbol{\pi}$ matrix. We have
\begin{eqnarray}
&\boldsymbol{\gamma}_{i}\boldsymbol{\gamma}_{j}\boldsymbol{\gamma}_{k}
\!=\!\boldsymbol{\gamma}_{i}\eta_{jk}\!-\!\boldsymbol{\gamma}_{j}\eta_{ik}
\!+\!\boldsymbol{\gamma}_{k}\eta_{ij}
\label{id2}
\end{eqnarray}
as a general identity. Defining $\overline{\psi}\!=\!\psi^{\dagger}\boldsymbol{\gamma}^{0}$ we can build the bi-linear spinor quantities as
\begin{eqnarray}
&\ \ \ \ \ \ \ \ \ \ \ \ \ \ \ \ \ \ \ \ \ \ \ \
U^{a}\!=\!\overline{\psi}\boldsymbol{\gamma}^{a}\psi\\
&\Theta\!=\!i\overline{\psi}\boldsymbol{\pi}\psi\ \ \ \
\ \ \ \ \ \ \ \ \Phi\!=\!\overline{\psi}\psi
\end{eqnarray}
which are all real tensors. They verify
\begin{eqnarray}
&U_{a}U^{a}\!=\!\Phi^{2}\!+\!\Theta^{2}
\end{eqnarray}
as Fierz identities. When $\Phi^{2}\!+\!\Theta^{2}\!\neq\!0$ we can always write the spinor in polar form, which is given, in a representation for which $\boldsymbol{\gamma}^{0}\!=\!\boldsymbol{\sigma}^{1}$ and $\boldsymbol{\gamma}^{1}\!=\!-i\boldsymbol{\sigma}^{2}$ (where $\boldsymbol{\sigma}^{1}$ and $\boldsymbol{\sigma}^{2}$ are two Pauli matrices), as
\begin{eqnarray}
&\psi\!=\!\phi\ e^{-\frac{i}{2}\beta\boldsymbol{\pi}}
\ \boldsymbol{L}^{-1}\left(\!\begin{tabular}{c}
$1$\\
$1$
\end{tabular}\!\right)
\label{spinor2}
\end{eqnarray}
for a pair of functions $\phi$ and $\beta$ and for some $\boldsymbol{L}$ with the structure of a spinor transformation. In this form
\begin{eqnarray}
&\Theta\!=\!2\phi^{2}\sin{\beta}\ \ \ \
\ \ \ \ \Phi\!=\!2\phi^{2}\cos{\beta}
\end{eqnarray}
so that $\phi$ and $\beta$ are a scalar and a pseudo-scalar. Then
\begin{eqnarray}
&\!\!U^{a}\!=\!2\phi^{2}u^{a}
\end{eqnarray}
is the velocity vector. It verifies
\begin{eqnarray}
&u_{a}u^{a}\!=\!1
\end{eqnarray}
as normalization condition. Therefore $\phi$ and $\beta$ are the $2$ variables that represent the two true degrees of freedom.

The $4$-dimensional construction of both tensorial connections can be re-done exactly also in the $2$-dimensional case. Yet, now the space-time tensorial connection is
\begin{eqnarray}
&R_{\alpha\beta\mu}\!=\!R_{\alpha}g_{\beta\mu}\!-\!R_{\beta}g_{\alpha\mu}
\label{R2d}
\end{eqnarray}
where $R_{\alpha}$ is a real vector. With it and $P_{\mu}$ we have
\begin{eqnarray}
&\!\!\!\!\!\!\!\!\boldsymbol{\nabla}_{\mu}\psi\!=\!(\nabla_{\mu}\ln{\phi}\mathbb{I}
\!-\!\frac{i}{2}\nabla_{\mu}\beta\boldsymbol{\pi}
\!-\!\frac{1}{2}R^{\alpha}\varepsilon_{\alpha\mu}\boldsymbol{\pi}
\!-\!iP_{\mu}\mathbb{I})\psi
\end{eqnarray}
where $2\boldsymbol{\sigma}_{ab}\!=\!\varepsilon_{ab}\boldsymbol{\pi}$ was used.

The Dirac field equation re-written as
\begin{eqnarray}
&\nabla_{\mu}\beta\!-\!2P^{\alpha}\varepsilon_{\alpha\mu}
\!+\!2mu^{\alpha}\varepsilon_{\alpha\mu}\cos{\beta}\!=\!0\\
&\nabla_{\mu}\ln{\phi^{2}}\!+\!R_{\mu}
\!+\!2mu^{\alpha}\varepsilon_{\alpha\mu}\sin{\beta}\!=\!0
\end{eqnarray}
shows that the gauge tensorial connection appears only in the field equation for the chiral angle while the space-time tensorial connection appears only in the field equation for the module. In this sense, they remain decoupled.

In $2$ dimensions, the Ricci curvatures are
\begin{eqnarray}
&R_{ab}\!=\!-\nabla_{i}R^{i}g_{ab}\\
&R\!=\!-2\nabla_{i}R^{i}
\end{eqnarray}
meaning that the Einstein tensor is zero, as expected. In the last relation we see that
\begin{eqnarray}
&-\frac{1}{2}R\!=\!\nabla_{\rho}G^{\rho}_{2}
\end{eqnarray}
for some vector
\begin{eqnarray}
&G^{\rho}_{2}\!=\!R^{\rho}
\end{eqnarray}
which is the Euler topological current.

Additionally, one can see that
\begin{eqnarray}
&\frac{1}{2}qF_{\alpha\nu}\varepsilon^{\alpha\nu}\!=\!\nabla_{\mu}K^{\mu}_{2}
\end{eqnarray}
for some axial-vector
\begin{eqnarray}
&\!\!\!\!K^{\mu}_{2}\!=\!P_{\nu}\varepsilon^{\nu\mu}
\end{eqnarray}
which is the Pontryagin topological current.

Once again we have written both Euler and Pontryagin topological currents in terms of the space-time and gauge tensorial connections, these last appearing as sources in-side the Dirac differential field equations.
\subsubsection{(0+2)-signature}
The $(0\!+\!2)$-dimensional case has Minkowski matrix in which the two elements have the same sign.

The Clifford and sigma matrices verify $\boldsymbol{\gamma}^{a}\boldsymbol{\pi}\!=\!i\varepsilon^{ab}\boldsymbol{\gamma}_{b}$ as well as $2i\boldsymbol{\sigma}_{ab}\!=\!\varepsilon_{ab}\boldsymbol{\pi}$ defining the $\boldsymbol{\pi}$ matrix. We have (\ref{id2}) still valid. However, the adjoint is now $\overline{\psi}\!=\!\psi^{\dagger}$ and hence the bi-linear spinor quantities are
\begin{eqnarray}
&\ \ \ \ \ \ \ \ \ \ \ \ \ \ \ \ \ \ \ \ \ \ \ \
S^{a}\!=\!\overline{\psi}\boldsymbol{\gamma}^{a}\psi\\
&\Theta\!=\!\overline{\psi}\boldsymbol{\pi}\psi\ \ \ \
\ \ \ \ \ \ \ \ \Phi\!=\!\overline{\psi}\psi
\end{eqnarray}
all being real tensors. They verify
\begin{eqnarray}
&S_{a}S^{a}\!=\!\Phi^{2}\!-\!\Theta^{2}
\end{eqnarray}
where $\Phi^{2}\!\geqslant\!\Theta^{2}$ since in this signature the norm of vectors is always positive. It is always possible to write the spinor in polar form, which is given, in the representation where $\boldsymbol{\gamma}^{i}\!=\!\boldsymbol{\sigma}^{i}$ (with $\boldsymbol{\sigma}^{i}$ two Pauli matrices), according to
\begin{eqnarray}
&\!\psi\!=\!\phi\ e^{-\frac{1}{2}\eta\boldsymbol{\pi}}
\boldsymbol{L}^{-1}\left(\!\begin{tabular}{c}
$1$\\
$1$
\end{tabular}\!\right)
\end{eqnarray}
for some $\phi$ and $\eta$ that are real. Then
\begin{eqnarray}
&\Theta\!=\!-2\phi^{2}\sinh{\eta}\ \ \ \
\ \ \ \ \Phi\!=\!2\phi^{2}\cosh{\eta}
\end{eqnarray}
so that $\phi$ and $\eta$ are a scalar and a pseudo-scalar. Also
\begin{eqnarray}
&S^{a}\!=\!2\phi^{2}s^{a}
\end{eqnarray}
is the spin axial-vector. It verifies
\begin{eqnarray}
&s_{a}s^{a}\!=\!1
\end{eqnarray}
as normalization. So $\phi$ and $\eta$ are the degrees of freedom of the system. Notice, however, that the change of signature entailed a corresponding change to hyperbolic functions.

The space tensorial connection given in \eqref{R2d} is the same as it does not depend on the signature. However now
\begin{eqnarray}
&\boldsymbol{\nabla}_{\mu}\psi\!=\!(\nabla_{\mu}\ln{\phi}\mathbb{I}
\!-\!\frac{1}{2}\nabla_{\mu}\eta\boldsymbol{\pi}
\!-\!R^{\alpha}\boldsymbol{\sigma}_{\alpha\mu}
\!-\!iP_{\mu}\mathbb{I})\psi
\end{eqnarray}
as covariant derivative.

The Dirac field equations are
\begin{eqnarray}
&\nabla_{k}\eta\!-\!2P^{b}\varepsilon_{bk}\!-\!2m\cosh{\eta}\varepsilon_{kb}s^{b}\!=\!0\\
&\nabla_{k}\ln{\phi^{2}}\!+\!R_{k}\!+\!2m\sinh{\eta}\varepsilon_{kb}s^{b}\!=\!0
\end{eqnarray}
still with the gauge and space tensorial connections that appear as two separate external sources.

The structure of the Ricci scalar is also unchanged, so that $G^{\rho}_{2}\!=\!R^{\rho}$ is still the Euler topological current.
\subsection{3-dimensions}
\subsubsection{Any-signature}
The $3$-dimensional case that we are going to consider is peculiar in many senses, one of which being that, regardless the signature, the polar decomposition always yields the same outcome. It is with no loss of generality that it is possible to take into account only one signature, which will be chosen as that of the pure space. The Minkowski matrix is the identity in three dimensions, and the Levi-Civita fully antisymmetric tensor has three indices.

Clifford and sigma matrices verify $2\boldsymbol{\sigma}^{ab}\!=\!i\varepsilon^{abc}\boldsymbol{\gamma}_{c}$ with no $\boldsymbol{\pi}$ matrix defined. We have now
\begin{eqnarray}
&\boldsymbol{\gamma}_{i}\boldsymbol{\gamma}_{j}\boldsymbol{\gamma}_{k}
\!=\!\boldsymbol{\gamma}_{i}\eta_{jk}\!-\!\boldsymbol{\gamma}_{j}\eta_{ik}
\!+\!\boldsymbol{\gamma}_{k}\eta_{ij}\!+\!i\varepsilon_{ijk}\mathbb{I}\label{id3}
\end{eqnarray}
identically. With $\overline{\psi}\!=\!\psi^{\dagger}$ the bi-linear spinors are
\begin{eqnarray}
&S^{a}\!=\!\overline{\psi}\boldsymbol{\gamma}^{a}\psi\label{vector}\\
&\Phi\!=\!\overline{\psi}\psi\label{scalar}
\end{eqnarray}
all being real tensors. They verify
\begin{eqnarray}
&S_{a}\boldsymbol{\gamma}^{a}\psi\!-\!\Phi\psi\!=\!0
\end{eqnarray}
and
\begin{eqnarray}
&S_{a}S^{a}\!=\!\Phi^{2}
\end{eqnarray}
as Fierz identities. It is always possible to write
\begin{eqnarray}
&\!\psi\!=\!\phi \boldsymbol{L}^{-1}\left(\!\begin{tabular}{c}
$1$\\
$0$
\end{tabular}\!\right)
\label{spinor3}
\end{eqnarray}
with $\phi$ a real function. In polar form
\begin{eqnarray}
&\Phi\!=\!\phi^{2}
\end{eqnarray}
showing that $\phi$ is a scalar. As usual
\begin{eqnarray}
&S^{a}\!=\!\phi^{2}s^{a}
\end{eqnarray}
is the spin vector. It verifies
\begin{eqnarray}
&s_{a}\boldsymbol{\gamma}^{a}\psi\!-\!\psi\!=\!0
\label{au}
\end{eqnarray}
and
\begin{eqnarray}
&s_{a}s^{a}\!=\!1
\label{normal}
\end{eqnarray}
as normalization. So $\phi$ is the unique degree of freedom.

As in $4$ and $2$, also in $3$ dimensions $R_{ab\nu}$ and $P_{\nu}$ have the same definition. They can be collected together into
\begin{eqnarray}
&R^{\alpha\nu}_{\phantom{\alpha\nu}\mu}\!+\!2P_{\mu}s_{\rho}\varepsilon^{\rho\alpha\nu}\!=\!
\varepsilon^{\alpha\nu\rho}M_{\rho\mu}\!=\!\Sigma^{\alpha\nu}_{\phantom{\alpha\nu}\mu}
\label{Sigma3}
\end{eqnarray}
being the hodge dual of each other. Then
\begin{eqnarray}
&\!\!\!\!\boldsymbol{\nabla}_{\mu}\psi\!=\!(\nabla_{\mu}\ln{\phi}\mathbb{I}
\!-\!\frac{1}{2}\Sigma_{\alpha\nu\mu}\boldsymbol{\sigma}^{\alpha\nu})\psi
\end{eqnarray}
in which $2\boldsymbol{\sigma}^{ab}\!=\!i\varepsilon^{abc}\boldsymbol{\gamma}_{c}$ and (\ref{au}) have been used.

With polar variables we have that
\begin{eqnarray}
&R^{ij}_{\phantom{ij}\mu\nu}\!+\!2qF_{\mu\nu}s_{k}\varepsilon^{kij}
\!=\!\Sigma^{ij}_{\phantom{ij}\mu\nu}
\end{eqnarray}
is the compact form of the curvature.

The Dirac equation is equivalent to
\begin{eqnarray}
&M^{\rho}_{\phantom{\rho}\rho}\!-\!2m\!=\!0\label{const3}\\
&\Sigma_{\mu\nu}^{\phantom{\mu\nu}\nu}\!+\!\nabla_{\mu}\ln{\phi^{2}}\!=\!0\label{eq3}
\end{eqnarray}
in which we see a peculiar occurrence. Because the spinor has a single degree of freedom, in $3$ dimensions there are $3$ differential equations determining the $3$ derivatives of the degree of freedom. The Dirac equations are $2$ complex, or $4$ real, conditions. Of these $4$ conditions, $3$ preserve the status of differential equations, and hence $1$ must become a constraint. This is what happened to equation (\ref{const3}).

There is, in $3$ dimensions, no characteristic class, which means that now there is a priori no way to have currents whose divergence is zero in flat spaces. Nevertheless, one can still define currents that are divergenceless, whether in flat spaces or not, according to
\begin{eqnarray}
&G^{\mu}_{3}\!=\!\varepsilon^{\mu\nu\sigma}\varepsilon^{\omega\alpha\eta}s_{\omega}
(\Sigma_{\alpha\eta\nu\sigma}\!+\!2s^{i}s^{j}\Sigma_{i\alpha\nu}\Sigma_{j\eta\sigma})
\end{eqnarray}
which is a type of Euler-like topological current. Without any curvature of the manifold we have
\begin{eqnarray}
&G^{\mu}_{3}\!=\!4qF_{\nu\sigma}\varepsilon^{\nu\sigma\mu}
\end{eqnarray}
as expected for the magnetic components. On the other hand, without electrodynamics we get
\begin{eqnarray}
&G^{\mu}_{3}\!=\!\varepsilon^{\mu\nu\sigma}\varepsilon^{\omega\alpha\eta}s_{\omega}
(R_{\alpha\eta\nu\sigma}\!+\!2s^{i}s^{j}R_{i\alpha\nu}R_{j\eta\sigma})
\label{G3}
\end{eqnarray}
recovering the results of \cite{Golkar:2014paa}.

We notice also that by taking the trace $R_{\alpha\nu\sigma}g^{\nu\sigma}\!=\!R_{\alpha}$ one can prove that
\begin{eqnarray}
&K^{\mu}_{3}\!=\!\varepsilon^{\mu\nu\alpha}\nabla_{\nu}R_{\alpha}
\end{eqnarray}
is divergenceless in curved spaces and as such it is a well-defined type of Pontryagin-like topological current.
\section{Conclusion}
In this work, we have shown through various examples in different dimensions how the Euler and the Pontryagin topological currents could be re-written with real tensors called tensorial connections, which were also shown to be found as sources in the Dirac differential field equations.

The possibility to link topological features of manifolds to dynamical characters of spinors defined on those manifolds is the idea at the center of fundamental results like the Atiyah-Singer index theorem \cite{AS}. We regard results like the one presented here as another (small) step toward clarifying the connections between spinors and topology.

\

\textbf{Funding and acknowledgements}. This work is carried out in the framework of activities of the INFN Research Project QGSKY and funded by Next Generation EU through the project ``Geometrical and Topological effects on Quantum Matter (GeTOnQuaM)''.

\

\textbf{Conflict of interest}. The Author declares no conflict.

\end{document}